\documentclass{article}
\usepackage{graphicx} 
\usepackage{hyperref}

\title{Solar and Wind Power Forecasting: A Comparative Review of LSTM, Random Forest, and XGBoost Models}
\author{
Afsaneh Mollasalehi\thanks{Faculty of Technology, Natural Sciences and Maritime Sciences, University of South Eastern Norway, Kongsberg, Norway (email: salehi.afsane23@gmail.com).}
\and
Armin Farhadi\thanks{School of Electrical and Computer Engineering, College of Engineering, University of Tehran (email: armin.farhadi@ut.ac.ir).}
}

\date{} 

\begin{document}

\maketitle
\begin{abstract}
Rising global energy demand from population growth raises concerns about the sustainability of fossil fuels. Consequently, the energy sector has increasingly transitioned to renewable energy sources like solar and wind, which are naturally abundant. However, the periodic and unpredictable nature of these resources pose significant challenges for power system reliability. Accurate forecasting is essential to ensure grid stability and optimize energy management. But due to the high variability in weather conditions which directly affected wind and solar energy, achieving precise predictions remains difficult. Advancements in Artificial Intelligence (AI), particularly in Machine Learning (ML) and Deep Learning (DL), offer promising solutions to improve forecasting accuracy. The study highlights three widely used algorithms for solar and wind energy prediction: Long Short-Term Memory (LSTM), Random Forest (RF), and Extreme Gradient Boosting (XGBoost). These models are capable of learning complex patterns from historical and environmental data, enabling more accurate forecasts and contributing to the enhanced efficiency and reliability of renewable energy systems. This review aims to provide an overview on RF, XGBoost, and LSTM by conducting a comparative analysis across three essential criteria: research prevalence, model complexity, and computational execution time.
\end{abstract}


\section{Introduction}
The global rise in energy need, driven by rising global population, urbanization, and technological development, has intensified concerns about the availability of traditional energy sources~\cite{ref1}. Also, higher gas emission is the result of increased consumption of these limited sources~\cite{ref2,ref3}. Moreover, climate change driven by our reliance on fossil fuels has led to increasingly severe weather patterns and a reduction in biodiversity~\cite{ref4}.

All these reasons have paved the way toward Renewable Energy Sources (RES) such as solar, wind, and hydro. Among the above-mentioned RES, solar and wind power have become increasingly popular. However, the intermittent and weather-dependent manner of these sources create challenges~\cite{ref5}. 

Wind energy production is highly affected by the natural fluctuations in wind patterns and the impact of geographical features. Solar energy is also influenced by climate change which makes their integration into power grids complicated. To address the variation in renewable energy production caused by weather conditions, it is crucial for power grid systems to conduct accurate energy generation forecasts. Accurate forecasting not only enhances the stability and reliability of the power grid but also allows for better resource planning, and reduced reliance on fossil fuel-based systems. By anticipating fluctuations in solar and wind generation, grid operators can make informed decisions about energy storage, load balancing, and demand response strategies, ultimately supporting a smoother integration of renewable energy into the overall energy mix~\cite{ref6}. 

Over the years, in order to find the complexities existed in renewable energy sources, a wide range of forecasting approaches have been developed. These methods contain physical, statistical, Machine Learning (ML), and hybrid models. Physical approaches, such as Numerical Weather Prediction (NWP), are widely used for short-term forecasts based on meteorological data. Statistical models, like Autoregressive Integrated Moving Average (ARIMA) and Seasonal ARIMA (SARIMA), are suitable for analyzing time series data under relatively stable conditions. ML models—such as neural networks (like Artificial Neural Networks (ANN) and Long Short-Term Memory (LSTM) networks) and tree-based methods (such as Extreme Gradient Boosting (XGBoost) and Light Gradient Boosting Machine (LightGBM))—are effective at capturing complex, non-linear patterns in large datasets~\cite{farhadi2025joint,ref7,ref8, ref18}. 

To deal with the challenges of predicting solar and wind energy, including constant changes, complex patterns, and dependence on time-related factors, this study uses three well-known ML and Deep Learning (DL) models. These models are LSTM, XGBoost, and Random Forest (RF). We chose these models specifically because they excel at managing time series data, especially when it's influenced by weather conditions. Their effectiveness in this area makes them a strong choice for accurate analysis and forecasting. LSTM is a special type of Recurrent Neural Network (RNN) that is designed to work with sequence data. Unlike regular RNNs, LSTM can remember important information from earlier time steps using a built-in memory system. This makes it ideal for tasks like forecasting solar or wind energy, where past conditions can strongly affect future values~\cite{ref9}. 

XGBoost is an improved version of the gradient boosting method and performs well at finding complex relationships between different variables, such as how temperature, humidity, and past energy output affect future energy production. Its regularization features help avoid overfitting, and it works well with large and noisy datasets, which are common in energy forecasting~\cite{ref10,ref11}.

RF is an ensemble method made up of many decision trees. While it does not directly model time-based patterns, it becomes effective for time series forecasting when we add features such as previous values or moving averages. RF is known for being stable and handling noisy or incomplete data well. This is useful in renewable energy forecasting, where weather data can be unpredictable or missing~\cite{ref12}.

\subsection{Application of ML in Renewable Energy Forecasting}
ML, as a subset of artificial intelligence, creates the potential of learning patterns in data and making forecasting without being explicitly programmed. In the case of time series forecasting, ML algorithms have outperformed typical statistical methods, offering better generalization and adaptability, particularly in the domains of renewable energy~\cite{ref7,ref8}.

In recent years, a wide range of ML approaches ranging from ensemble methods, DL, and hybrid techniques have been applied to forecast solar and wind power generation. These methods utilize weather data, temporal features, and environmental factors to forecast short-term and long-term energy outputs. Studies have shown that methods like RF, XGBoost, and LSTM effectively capture the variability and nonlinear behavior of renewable energy systems~\cite{ref12,ref13}.

\subsection{Model Description and Working Principles of LSTM, XGBoost, and RF Models}
This section delivers an overview of the key concepts in LSTM, XGBoost, and RF models which are essential to have a better understanding of the methods utilized in renewable energy forecasting. This part also aims to demonstrate how LSTM, XGBoost, and RF algorithms effectively forecast energy production from renewable sources like solar and wind power.

\subsubsection{LSTM}
LSTM networks have become essential tools for modeling time series data, particularly in the energy forecasting domain. The LSTM model has the ability to capture time-related patterns in time series problems. Unlike standard RNNs, which struggle with the vanishing gradient problem during training, LSTM tackles this issue through the use of memory cells that help maintain information across long sequences. 

\begin{figure}[t!]
    \centering
    \includegraphics[width=0.45\textwidth]{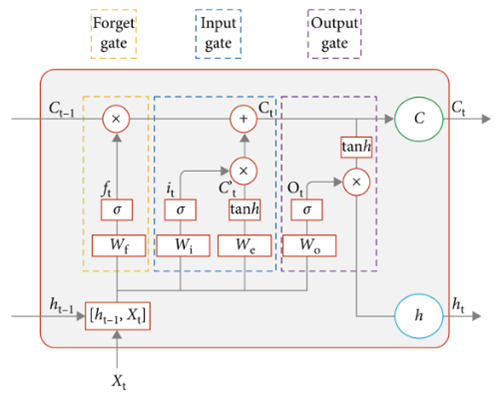}
    \caption{LSTM memory block diagram~\cite{ref14}.}
    \label{fig:lstm_block}
\end{figure}

At the core of the LSTM architecture is a sophisticated gating mechanism composed of input, forget, and output gates. These gates work together to regulate the flow of information, determining what to retain, update, or discard at each time step. As illustrated in Fig.~\ref{fig:lstm_block}, when the input gate is activated, new data enters the memory cell. If needed, the forget gate removes outdated information, while the output gate decides whether the current memory state should be passed forward. Together, these components ensure efficient information flow and help the network focus on the most relevant signals in the data. This configuration enables LSTMs to retain important context over extended periods—a critical feature when forecasting energy generation, since weather conditions can significantly influence future outcomes~\cite{ref14}. 

Additionally, LSTMs excel at capturing nonlinear relationships in data, making them well-suited for understanding how multiple factors such as temperature, wind speed, sunlight, and seasonal patterns interact to affect energy output. Given that weather data is inherently time-based and consists of continuously recorded variables, it fits naturally into the framework of time series analysis. These variables often display patterns such as seasonal cycles, long-term trends, and sudden fluctuations, all of which are crucial for accurate forecasting~\cite{ref10,ref11}.


LSTM is an appropriate solution for renewable energy forecasting because it is practically designed to handle time-dependent data such as weather patterns, which are substantial to forecasting solar and wind energy. LSTM can remember long-term trends because of its memory cells and gating system. It also handles complex, nonlinear relationships between weather features and the time-series nature of renewable energy data, making it suitable for the task of wind and solar energy forecasting~\cite{ref14}.

\subsubsection{XGBoost}
The gradient boosting is a method used in the XGBoost model. A sequence of independent decision trees is made in this method. Each new tree is designed in a way to improve the mistakes on the previous tree. In this way, the complexities and non-linear behavior of renewable energy sources can be detected. Moreover, this algorithm uses a regularization term in the loss function to avoid overfitting. 

\begin{figure}[ht]
    \centering
    \includegraphics[width=0.45\textwidth]{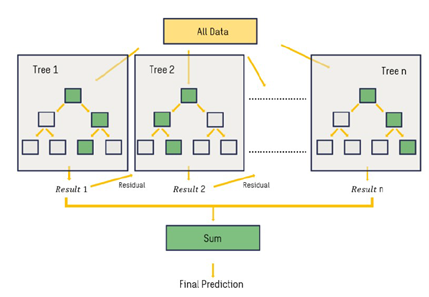}
    \caption{A diagram of the XGBoost algorithm~\cite{ref15}.}
    \label{fig:xgboost}
\end{figure}

As illustrated in Fig.~\ref{fig:xgboost}, XGBoost is well-suited for solar and wind energy prediction because it can handle complex, nonlinear patterns in the data by building a series of decision trees that learn from previous mistakes. This makes it effective at modeling the unpredictable nature of renewable energy. It also includes a regularization term to reduce overfitting, which helps the model stay accurate even when dealing with many influencing factors like weather and time~\cite{ref10,ref11}. 

\subsubsection{RF}
\begin{figure}[b!]
    \centering
    \includegraphics[width=0.45\textwidth]{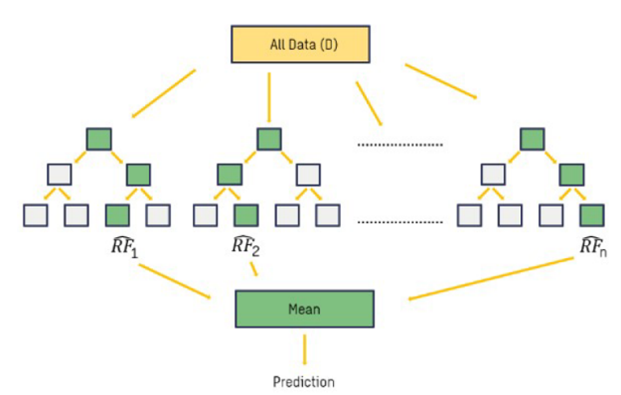}
    \caption{The structure of the RF model~\cite{ref15}.}
    \label{fig:rf}
\end{figure}

RF is a tree-based model using bootstrapping and bagging techniques to generate random combinations among trees and improve accuracy~\cite{ref16}. As illustrated in Fig.~\ref{fig:rf}, all data is located in box ``D'' on the top of the tree. The training data is randomly separated into smaller trees where $\hat{RF_i}$ indicates the predicted values of each tree $i$. Moreover, only a few randomly selected features are utilized for each part. This randomness helps make the model more diverse and accurate. At each green box, decisions are made. Then RF predictions are finalized by averaging all individual trees~\cite{ref17}.

Unlike some other models, the trees in RF are allowed to grow deep, without manually setting strict stopping rules. Instead, stopping points are usually determined automatically using metrics like the Gini index, RMSE, or MSE. In the end, only the trees that make good predictions are kept in the final model—those that perform poorly are left out~\cite{ref9}.

RF is an appropriate option for forecasting solar and wind energy. This model constructs multiple decision trees using random selection subsets of input data which improve the model’s robustness. RF also reduces overfitting by averaging the predictions from all these trees. RF effectively manages noisy and complex data, making it ideal for the unpredictable nature of renewable energy sources. Additionally, this model automatically selects the best-performing trees resulting in a reliable final model that requires minimal manual tuning~\cite{ref12}.

\subsection{Comparison of RF, XGBoost, and LSTM in Renewable Energy Forecasting}
\begin{figure}[b!]
    \centering
    \includegraphics[width=0.45\textwidth]{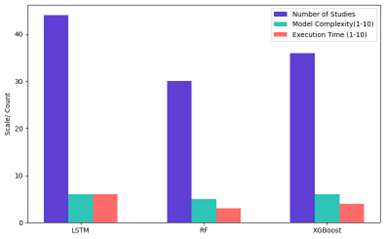}
    \caption{Comparison of RF, XGBoost, and LSTM models for renewable energy forecasting~\cite{ref12}.}
    \label{fig:comparison}
\end{figure}

The bar chart in Fig.~\ref{fig:comparison} describes a comparative analysis of three ML models including LSTM, RF, and XGBoost based on three key criteria: Research Frequency, Model Complexity, and Execution Time.

\begin{enumerate}
    \item \textbf{Research Frequency:} reflects the level of academic attention each model has received. A higher value indicates greater adoption and validation in scientific literature. DL models, particularly LSTM, show the highest adoption, highlighting their perceived effectiveness in renewable energy prediction tasks~\cite{ref12}.
    
    \item \textbf{Model Complexity:} is rated from 1 to 10, indicating the level of computational complexity of each model. Tree-based methods like RF and XGBoost exhibit lower complexity, while LSTM requires more resources due to its deep architecture~\cite{ref12}.
    
    \item \textbf{Execution Time:} is also on a 1–10 scale and measures how computationally intensive a model is during training. Simpler models execute faster and work better for real-time applications, while LSTM is slower but often chosen when high accuracy is more important~\cite{ref12}.
\end{enumerate}

\subsection{Result of the Analysis}
The comparison reveals the following perspectives:

\begin{itemize}
    \item Ensemble techniques such as RF and XGBoost offer an effective balance between predictive accuracy and computational efficiency, contributing to their widespread adoption in renewable energy forecasting~\cite{ref12}.
    
    \item DL models, especially LSTMs, provide superior forecasting performance. However, this comes with increased computational time and longer processing steps~\cite{ref12}.
\end{itemize}

This comparative evaluation offers a systematic framework for selecting suitable ML models based on the trade-offs among accuracy, model complexity, and execution time in renewable energy forecasting contexts~\cite{ref12}.

\section{Conclusion}
The energy systems have experienced a significant shift toward the adoption of ML and DL techniques. As fossil fuel reserves continue to diminish and global energy demand rises, the integration of RES, particularly wind and solar, has become increasingly important. However, the variable and weather-dependent nature of these sources introduces considerable uncertainty into energy supply. To manage this variability and maintain grid functionality, accurate forecasting models are essential. ML provides powerful tools for modeling the complicated, nonlinear, and dynamic patterns inherent in the production of renewable energy. Although a variety of approaches have been explored for this purpose, only a few have demonstrated consistently strong performance. Among these, RF, XGBoost, and LSTM networks have shown particularly promising results. These models offer satisfactory advantages in terms of handling noise and learning temporal patterns, making them appropriate for forecasting applications in the energy industry. In future research, these algorithms can be applied to suitable datasets for further evaluation and assessment of their effectiveness in renewable energy forecasting. Such investigations would help prove the practical usage of RF, XGBoost, and LSTM models and uncover new insights to improve prediction accuracy and system reliability in varying renewable energy contexts.

\bibliographystyle{IEEEtran}
\bibliography{Bibliography1}

\end{document}